# Paraxial micro earthquake: a natural effective multi-purpose check shot for downhole earthquake monitoring

*Ruiqing He and Björn Paulsson*

*Paulsson, Inc., 16543 Arminta Street, Van Nuys, California 91406, USA*

## Abstract

Downhole earthquake monitoring, without the effects from the overburden, can record better seismic data than monitoring on surface. However, in order to reasonably use the downhole vector seismic data, a constant challenge is how to accurately orient the downhole radial-component seismometers. A common practice is to use offset check shots on or near the surface. However, in areas with complex geologies, this routine may result in significant orientation errors. A ParAxial Micro Earthquake (PAME) is a micro earthquake at a close distance to the seismometers and near the extended path of the borehole's trajectory. It is rarely recorded during downhole earthquake monitoring unless designed for. If it is recorded, it can be a real treasure not only for P-wave and S-wave velocities' profiling, but for the downhole seismometers' orientation. As an example, during April to May in 2005, Paulsson installed an 80-level 3-component VSP (Vertical Seismic Profiling) array in the SAFOD (San Andreas Fault Observatory at Depth) main hole at Parkfield, California, and continuously recorded seismic data for about 13 days. Large charge offset check shots at 13 different locations near the surface were detonated in order to orient the downhole geophones; the orientation results were unsatisfactory but went unnoticed or unsolved. Besides this, a small charge "zero-offset" check shot was detonated near the wellhead in order to get the P-wave and S-wave velocity profiles, but only the P-wave velocity profiling was successful. Fortunately, we recorded a few PAMEs, through which we not only obtained better P-wave and S-wave velocity profiles, but satisfactorily oriented the downhole geophones.

## Introduction

Downhole earthquake monitoring, because of closer source-to-receiver distance, and avoiding the notorious near-surface complexities, can record better seismic data than monitoring on surface. Certainly, this advantage is at the cost of more expensive deployment. For this reason, downhole earthquake monitoring is not as commonly used as surface seismic monitoring. However, in some crucial areas, downhole earthquake monitoring is necessary.

If multi-level downhole seismometers are used, it is also known as VSP (Vertical Seismic Profiling), although it can also be deployed into deviated or horizontal boreholes. The downhole seismometers can be single component, for instance hydrophone or Distributed Acoustic Sensing (DAS), but in many cases, orthogonal 3C (3-Component) seismometers are used. When the borehole is vertical, the axial-component seismometers always point in the vertical direction, while the other two radial components point to some unknown horizontal directions. This is because, after deployment, the wireline (cable or tubing) on which the seismometers are attached has twisted. In the cases of deviated or horizontal boreholes, the axial-component seismometers always point in the borehole's trajectory direction, and the radial-component seismometers' directions are still unknown.

Before the data recorded by the radial-component seismometers can be reasonably used, these seismometers' directions must be found out. Ideally, auxiliary orientation instruments, such as inclinometers, magnetometers or gyros can be installed along with the seismometers in the borehole (Naville et al., 2017), but this costs significantly more. For this reason, this method is rarely used, although we have seen the necessity to use it more in the future.

A more common (if not the most common) practice is to use remote check shots and rotate the recorded radial-component direct P-wave waveform, so that it is maximized on one of the radial components (Greenhalgh and Mason, 1995). This method assumes most of the direct P-wave energy travels in a known direction in the two radial dimensions. The optimal location for the check shot is underground at the side of the downhole seismometers to avoid the bending of seismic rays in the overburden. This will also significantly increase the cost unless there is an existing nearby borehole ready for such use, like in hydraulic fracking.

In many cases, the check shots are carried out on or near the surface at an offset (a considerable distance away from the borehole). We know that the near surface almost always bends seismic rays in the vertical direction because of the ubiquitous vertical velocity heterogeneity. But if the horizontal velocity heterogeneity is not strong, so that most of the direct P-wave energy travels within the vertical plane from the source to the receiver, it is still possible to orient the radial-component seismometers.

In areas with complex geologies, people realize that the overburden's horizontal velocity heterogeneity may cause significant orientation errors (Van Dok et al. 2016). In such cases, normally a 3D velocity model is sought to mitigate this problem. Building a subsurface 3D velocity model needs a lot of work and its accuracy is still difficult to verify.

Besides the above active check shot methods, which are specially designed for the downhole seismometers' orientation, sometimes the recorded downhole seismic data contain information to orient the seismometers. Scholz et al. (2017), oriented ocean-bottom (similar to downhole) seismometers from P-wave and Rayleigh wave polarizations. Ensing and Wijk (2019) estimated the orientation of borehole seismometers from ambient seismic noise. Park et al. (2022) oriented borehole seismometers using ubiquitously present Rayleigh waves. The accuracy of these passive methods depends on the extent of the knowledge about the seismic waves being used and the geology they have traveled.

A ParAxial Micro Earthquake (PAME) is a micro earthquake at a close distance to the seismometers and near the extended path of the borehole's trajectory. It is rarely recorded during downhole earthquake monitoring unless people properly plan the acquisition to record it. If it is recorded, it can be a truly zero-offset check shot for better P-wave and S-wave velocities' profiling than the active "zero-offset" check shots on the surface. If we also know the shearing direction of the PAME, it can also be a better way to orient the downhole seismometers than the active offset check shots on the surface.

# PAME

The inaccuracy of remote methods in determining the downhole seismometers' directions mainly roots in people's uncertainty about the geology, especially the notorious near surface, through which the seismic waves have traveled. The optimal position for the check shots is as close to the seismometers to be oriented as possible, so that the unknown geology's effect is minimized. A PAME is a micro earthquake at a close distance from the downhole seismometers which makes it possibly an ideal check shot.

Natural earthquakes usually are characterized by strong shear waves. Because a PAME is at a position near the extended path of the borehole's trajectory, the downhole axial-component seismometers will record very weak S waves. On the other hand, the radial-component seismometers will record strong S waves, but very weak P waves as compared with the S waves. This exactly is the feature applicable to identify a PAME.

The module of the PAME's 3C seismic data usually has both clear P waves and S waves which makes it an ideal "zero-offset" check shot to obtain both the P-wave and S-wave velocity profiles along the borehole. In the case of deviated or horizontal boreholes, a PAME can be better than an active "zero-offset" check shot on the surface, because the latter actually has varying offsets to the downhole seismometers, while a PAME can be truly zero-offset.

Although being "zero-offset", sometimes a PAME can be a substitute of an offset check shot to orient the downhole radial-component seismometers. This is done by rotating its recorded direct S wave on the two radial components and maximizing it onto one of them. Now, the radial component with the maximum S wave has been oriented to the shearing direction of the PAME. Usually, we still do not know this direction. In such cases, further investigations may or may not help figure it out. Fortunately, in some cases, we do have a good knowledge about the PAME's shearing direction, like in the following case study.

# Case Study

SAFOD (San Andreas Fault Observatory at Depth) is a deep (about 3 kilometers) borehole observatory with different instruments deployed to study the earthquake mechanism of the San Andreas fault at Parkfield, California, funded by the National Science Foundations (NSF) of the United States. During April-May in 2005, Paulsson installed a deployable 80-level 3C VSP array in the SAFOD main hole (a deviated well), and continuously recorded seismic data for about 13 days (with a few breaks).

Figure 1 shows the locations of the SAFOD site and offset check shots at 13 different locations. In the 3D model, the sparse dots represent the previous settings of the VSP array, and the dense dots represent the final positions of the 80-level 3C geophones. The deepest geophone was at 9,000 ft measured depth. Small charges of 5 lb. Pentolite at 10 ft depth were detonated as "zero-offset" check shots near the wellhead. Large charges of 80 lb. Pentolite at 100 ft depth were detonated as offset check shots.

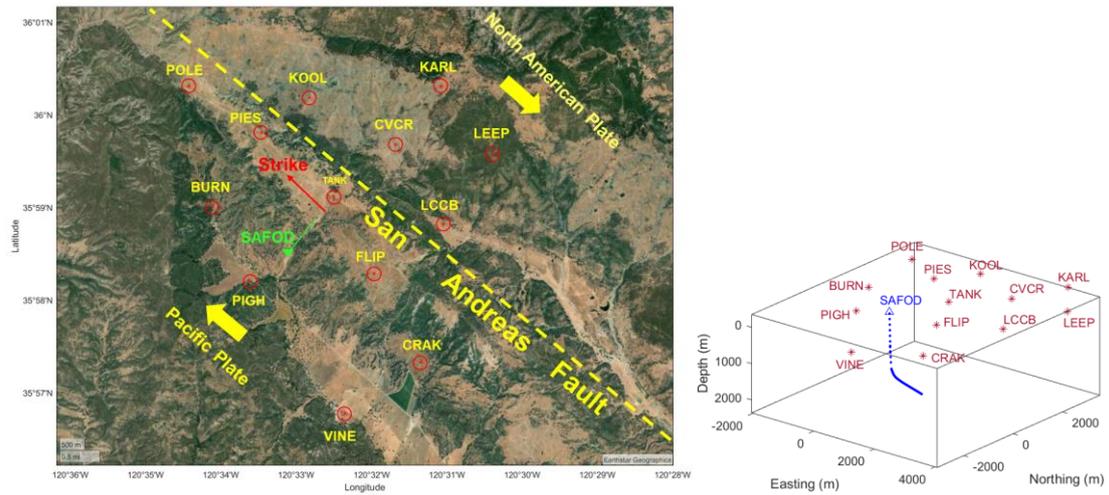

Fig. 1: The map of the 13 offset check shots around the SAFOD site (left), and a 3D illustration of the offset check shots and the SAFOD main hole, displayed with equal scales in the easting, northing, and vertical directions (right).

Figure 2 shows a geological model at the SAFOD site. It shows strong heterogeneities in both vertical and horizontal directions. The Axial direction is shown as the upward borehole direction. The unknown-direction radial-component geophones will be rotated into the Planar and Strike directions. In here, the Planar direction is defined as the direction downward perpendicular to the Axial direction and within the vertical plane. The Strike direction is defined as the inward direction orthogonal to both the Axial and Planar directions.

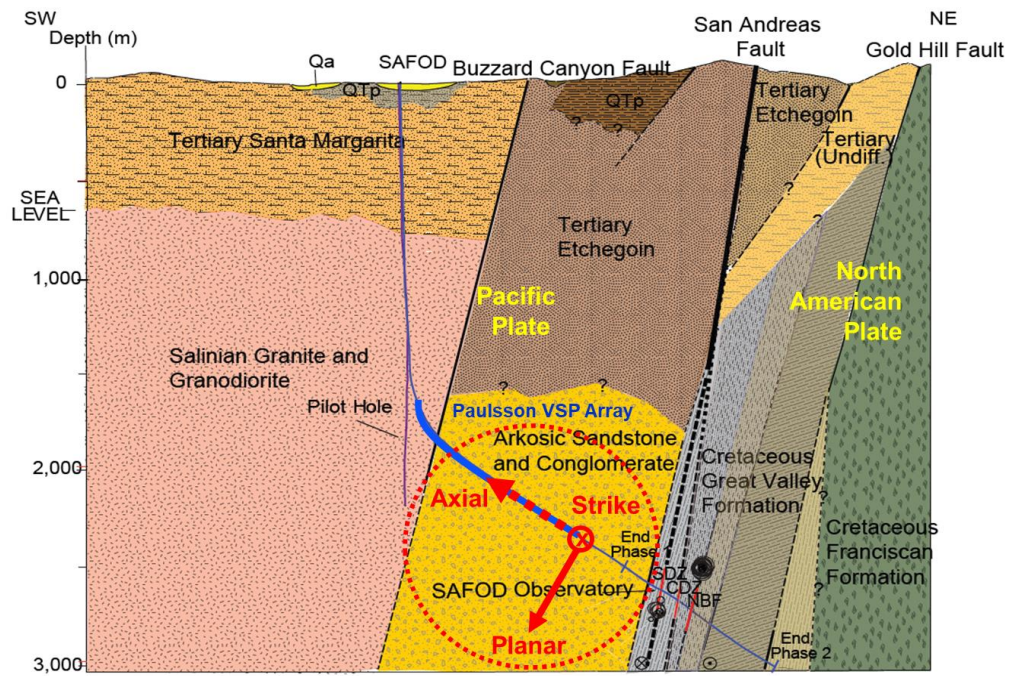

Fig. 2: A geological model at the SAFOD site (modified after Carpenter et al., 2011). In this paper, the two radial-component geophones will be rotated into the Planar and Strike directions as shown.

On the same well pad, SAFOD has another vertical well called the pilot hole, in which Duke University has installed a 32-level 3C VSP array and recorded seismic data for a longer term. From the data, Oye and Ellsworth (2005) observed strong seismic ray bending and used a 3D velocity model to improve the average directional residual from 13º to 7º.

The Paulsson recorded VSP data in the main hole has been used by Chavarria et al. (2007) for fault zone characterization, and by Rentsch et al. (2010) to locate about 2 dozen micro earthquake events. The authors also oriented the downhole geophones using the offset check shots to locate 100 micro earthquake events and later found out that this orientation had significant errors.

Figure 3 shows the 3C seismic data of the "zero-offset" check shot (the deepest 3C geophones are in pod #1, and the highest 3C geophones are in pod #80). Although the direct P wave is clear on the axial component, but very noisy on the two radial components. So, this check shot is not suitable for orienting the radial-component geophones. Besides this, the direct S wave is not clearly present, so it is also difficult to perform the S-wave velocity profiling.

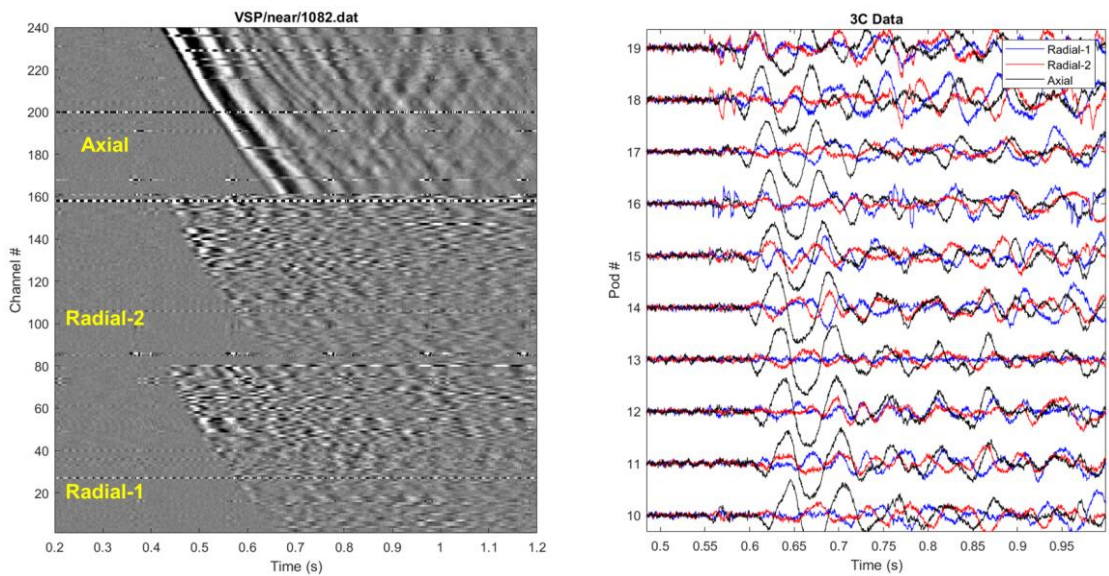

Fig. 3: The 3C seismic data of the "zero-offset" check shot (left) and some of its selected direct P-wave waveforms (right).

Figure 4 shows the 3C seismic data of the check shot TANK. It has high signal-to-noise (S/N) ratio, but the direct P-waves at the bottom levels seem to have been refracted, which makes it unideal for orienting the radial-component geophones.

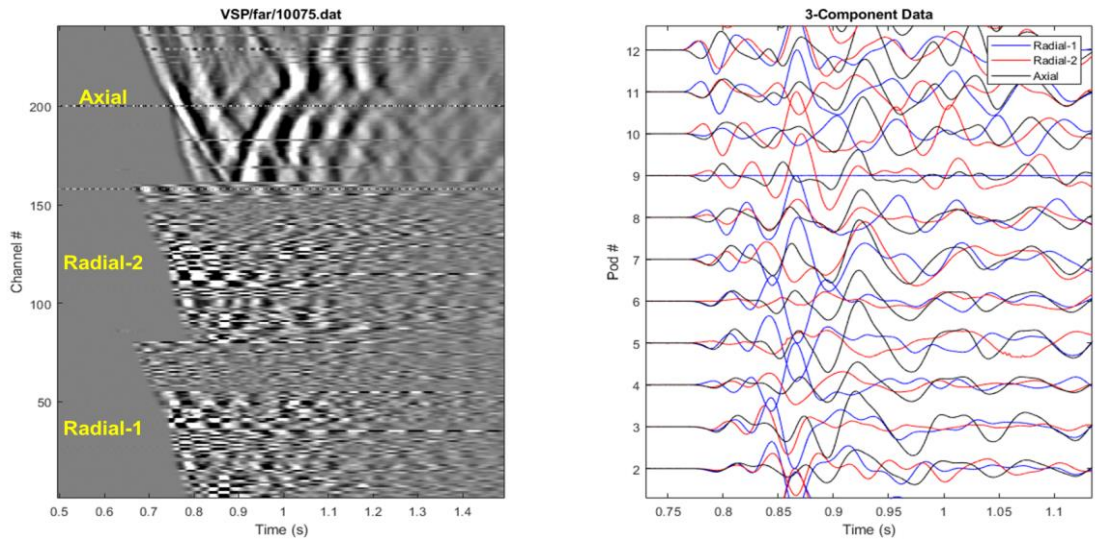

Fig. 4: The 3C seismic data of the check shot TANK (left) and some of its selected direct P-wave waveforms (right).

Figure 5 shows the 3C seismic data of the check shot PIGH. It also has a high S/N ratio, and the direct P-wave does not appear to have been refracted. However, the direct P wave on the two radial components is weak as compared with that on the axial component, which could result in unstable orientations of the radial-component geophones.

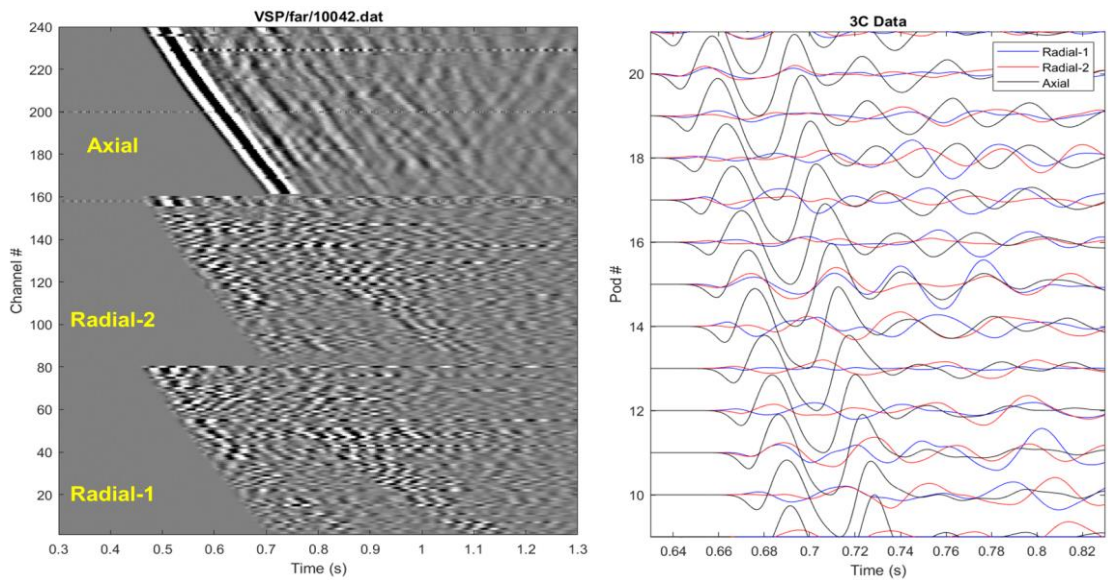

Fig. 5: The 3C seismic data of the check shot PIGH (left) and some of its selected direct P-wave waveforms (right).

The seismic data of the check shot BURN has a similar S/N ratio as that of PIGH. But it is unideal for the downhole geophones' orientation because the unknown vertical bending of the direct P wave will affect its direction in the two radial dimensions. The seismic data of the check shot FLIP has the same problem, in addition to an apparently refracted direct P wave. Other check shots are even more unideal for the downhole geophones' orientation, either because of their unsuitable locations, low S/N ratios, or apparently refracted direct P waves.

Fortunately, we recorded a few PAMEs. Figure 6 shows one of them that is used to orient the downhole geophones. Its direct P wave shows strong on the axial component but weak on the two radial components. On the other hand, its direct S wave shows strong on the two radial axial components but weak on the axial component. From the difference between the P-wave and S-wave arrival times, it is estimated that this PAME was about 530 m away (about 1/5 of the minimum distance to the surface check shots) and near the extended path of the borehole's trajectory. Figure 7 shows its rotated data with the direct S wave maximized on one of the radial components. This maximum S-wave direction is believed to be the Strike direction shown in figure 2, because we have good knowledge from long-year experiences that the San Andreas fault is a right-lateral strike-slip fault.

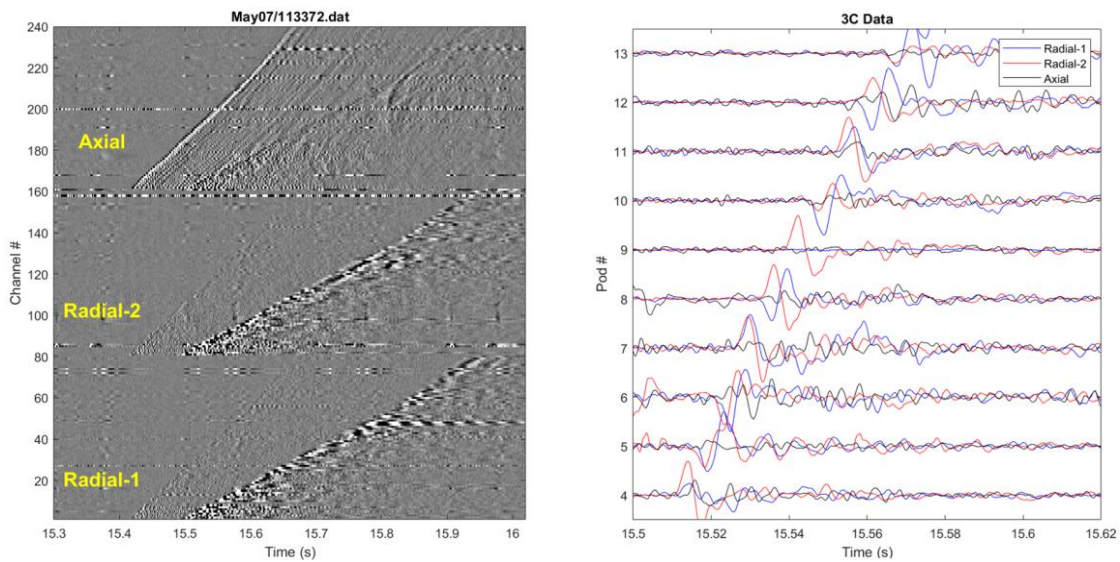

Fig. 6: The 3C seismic data of a PAME (left) and some of its selected direct S-wave waveforms (right).

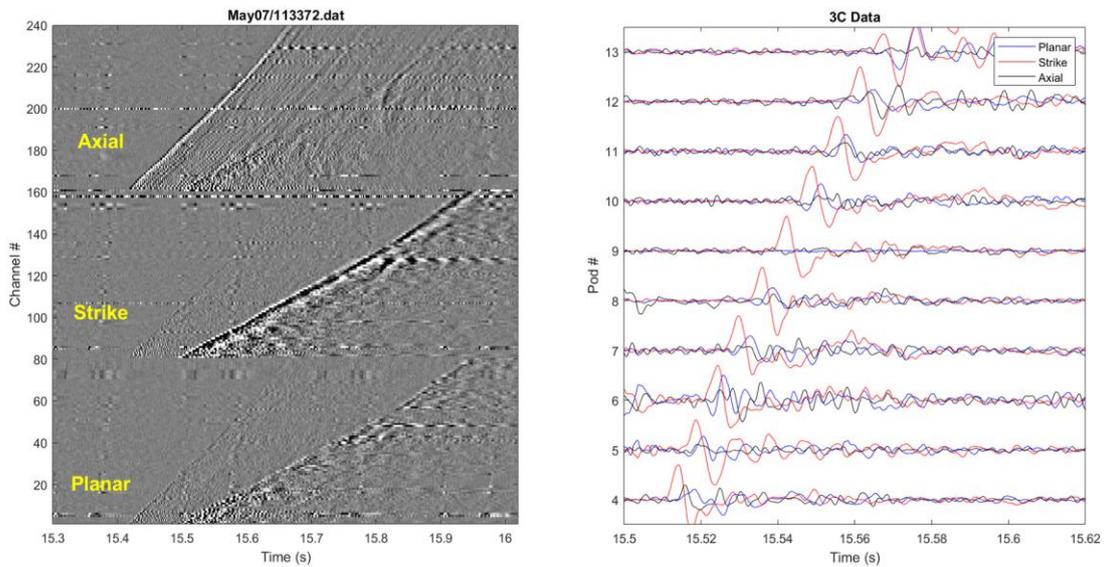

Fig. 7: The rotated seismic data (left) and some of its selected direct S-wave waveforms (right) of the PAME shown in figure 6.

After the downhole radial-component geophones have been oriented by the PAME's S wave, figure 8 shows the rotated seismic data of the check shot PIGH. Surprisingly the direct P wave shows stronger on the Strike component, instead of on the Planar component as expected if the horizontal velocity heterogeneity were weak. Figure 9 shows the relative bearing angles (the clockwise angle away from the upward planar direction when looking downward along the borehole) of the direct P wave in the rotated seismic data. Besides fluctuating, the relative bearing angles show an average of 94°, which means the direct P wave was almost horizontally hitting the VSP array from the west. In the past, orienting the downhole geophones using this check shot had caused tremendous errors.

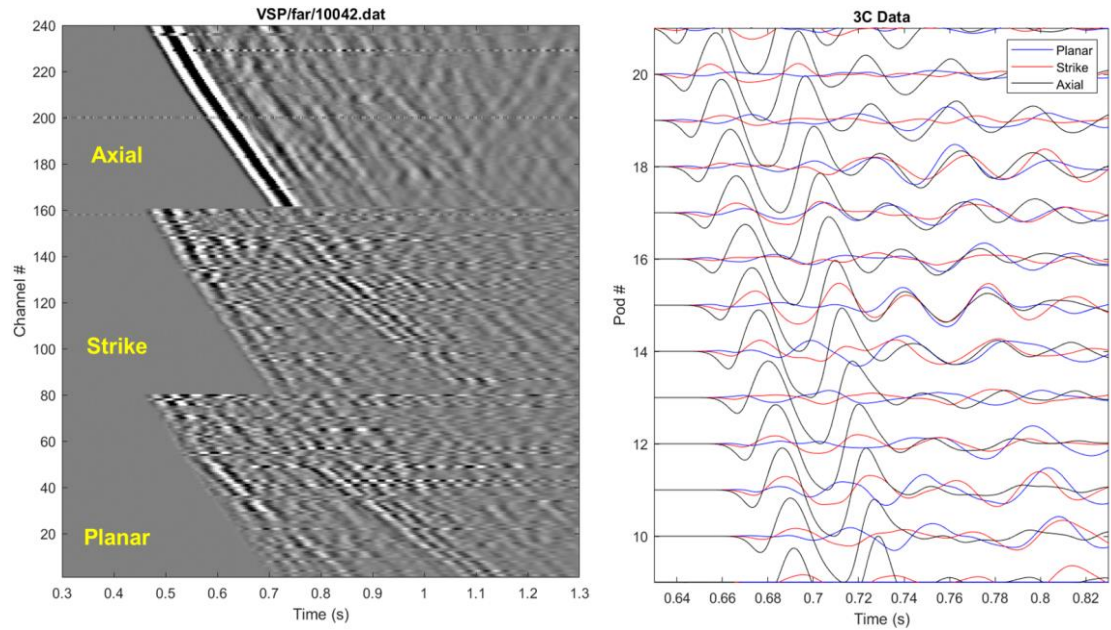

Fig. 8: The rotated seismic data (left) and some of its selected direct S-wave waveforms (right) of the check shot PIGH shown in figure 5.

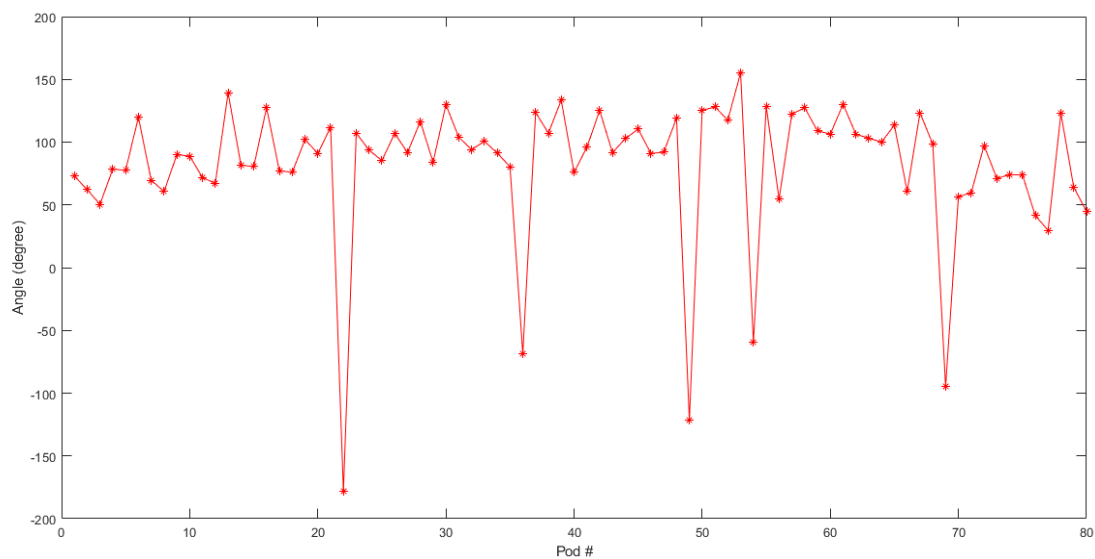

Fig. 9: The relative bearing angles of the direct P wave at the 80-level VSP geophones in the check shot PIGH.

Besides for the downhole radial-component geophones' orientation, the same PAME was used for P-wave and S-wave velocities' profiling. Figure 10 shows the comparisons of VSP P-wave and S-wave velocity profiles vs sonic logs. The result from the "zero-offset" VSP contains three settings of the VSP array at different depths. The P-wave velocity profiling using the PAME fits the sonic log better than using the surface "zero-offset" check shot. The S-wave velocity profile using the PAME also fits the shear sonic log nicely, while the surface "zero-offset" check shot failed to provide the S-wave velocity profile.

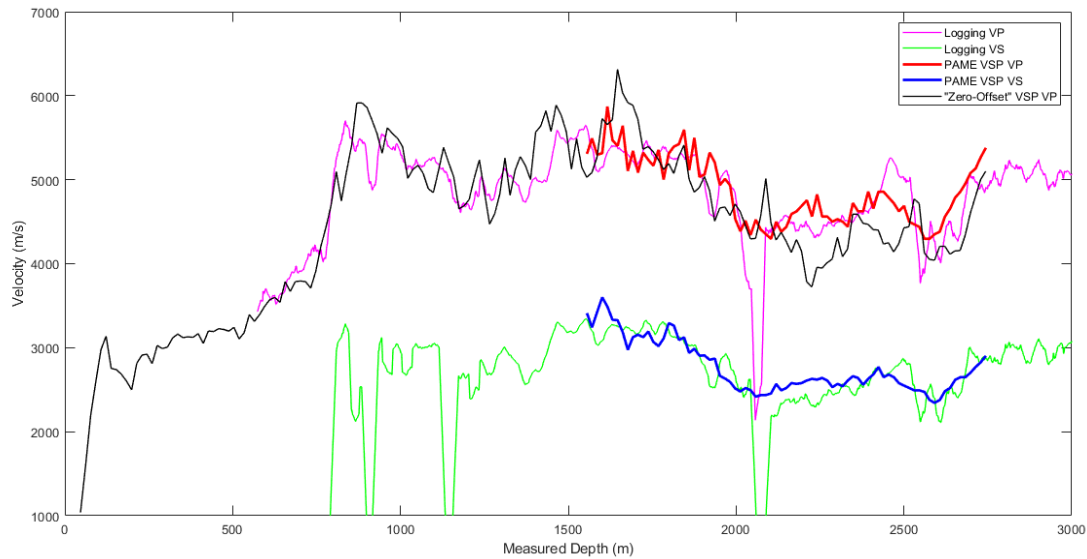

Fig. 10: VSP P-wave and S-wave velocity profiles using the surface "zero-offset" check shot (P-wave only) and the PAME (both P-wave and S-wave) vs sonic logs.

## Discussions

Seismometers' directional information is a prerequisite for many seismological analyses, and critically important for micro earthquakes' locating. However, seismic sensors including both borehole and surface seismometers are often installed in incorrect orientations (Park and Hong, 2024). The remote methods to orient the downhole seismometers heavily depend on the knowledge about the seismic waves being used and the geology they have experienced. In areas with complex geologies, it is also difficult to determine the severity of the orientation errors unless there is a comparison with the orientation result from a better way.

The problem roots in the lack of an accurate 3D model of the subsurface. In the above case study, we have full control and knowledge of the check shots used on the surface, but very little knowledge of the geology the seismic waves have traveled through. On the other hand, we have little knowledge about the PAME and the geology it has traveled. It is a rational choice rather than a science to decide which one should be used to orient the downhole geophones. In this case, we believe the uncertainty about geology is more severe, and should use the closer one to minimize the geology's effect. Indeed, in this case, micro earthquakes' locating using the geophone orientation by the PAME will be much more reasonable than using the geophone orientation by the offset check shots on the surface.

Besides its potential in orienting the downhole seismometers, and providing outstanding P-wave and S-wave velocity profiles, PAME is free of additional cost to record. For existing downhole earthquake monitoring, we should pay attention to look for PAMEs, and for future downhole earthquake monitoring, we should design the acquisition system to record PAMEs. This may reduce the need to design and execute active "zero-offset" and offset check shots.

## Acknowledgements

We thank NSF and other sponsors for supporting the Parkfield earthquake experiments, especially the SAFOD project.